\documentclass[12pt,smallextended,natbib,runningheads,epsfig]{article}
\usepackage{latexsym, amsfonts}
\usepackage{amssymb, amsmath}
\usepackage{amsfonts}
\usepackage{amsfonts}
\usepackage{graphicx}
\usepackage{adjustbox}
\usepackage{setspace}
\usepackage{rotating}
\usepackage{tikz}
\usepackage{bigints}
\usepackage{bm}
\usepackage{multirow}
\usepackage{changepage}  % For adjustwidth
\usepackage{subcaption}

\newcommand{\be}{\begin{equation}}
\newcommand{\ee}{\end{equation}}
\newcommand{\bea}{\begin{eqnarray}}
\newcommand{\eea}{\end{eqnarray}}
\newcommand{\bean}{\begin{eqnarray*}}
\newcommand{\eean}{\end{eqnarray*}}
\newcommand{\brray}{\begin{array}}
\newcommand{\erray}{\end{array}}
\newcommand{\ben}{\begin{equation}{nonumber}}
\newcommand{\een}{\end{equation}{nonumber}}

\newtheorem{dfn}{Definition}[section]
\newtheorem{thm}[dfn]{Theorem}
\newtheorem{lmma}[dfn]{Lemma}
\newtheorem{ppsn}[dfn]{Proposition}
\newtheorem{crlre}[dfn]{Corollary}
\newtheorem{xmpl}[dfn]{Example}
\newtheorem{rmrk}[dfn]{Remark}

\newcommand{\bdfn}{\begin{dfn}}
\newcommand{\bthm}{\begin{thm}}
\newcommand{\blmma}{\begin{lmma}}
\newcommand{\bppsn}{\begin{ppsn}}
\newcommand{\bcrlre}{\begin{crlre}}
\newcommand{\bxmpl}{\begin{xmpl}}
\newcommand{\brmrk}{\begin{rmrk}}
\newcommand{\edfn}{\end{dfn}}
\newcommand{\ethm}{\end{thm}}
\newcommand{\elmma}{\end{lmma}}
\newcommand{\eppsn}{\end{ppsn}}
\newcommand{\ecrlre}{\end{crlre}}
\newcommand{\exmpl}{\end{xmpl}}
\newcommand{\ermrk}{\end{rmrk}}

\def\a*{{\cal A}_{h,*}}
\def\B{{\cal B}(h)}
\def\B1{{\cal B}_1(h)}
\def\b{{\cal B}^{\rm s.a.}(h)}
\def\b1{{\cal B}^{\rm s.a.}_1(h)}

\numberwithin{equation}{section}

\begin{document}
\begin{center}
{\Large {\bf Bayesian Hybrid Machine Learning of Gallstone Risk}}\\[1.5ex]
{\large Chitradipa Chakraborty\textsuperscript{1,*} and Nayana Mukherjee\textsuperscript{2,$\dagger$}}\\
\end{center}

\begin{adjustwidth}{-0.5cm}{-0.5cm} 
\begin{abstract}
Gallstone disease is a complex, multifactorial condition with significant global health burdens. Identifying underlying risk factors and their interactions is crucial for early diagnosis, targeted prevention, and effective clinical management. Although logistic regression remains a standard tool for assessing associations between predictors and gallstone status, it often underperforms in high-dimensional settings and may fail to capture intricate relationships among variables. To address these limitations, we propose a hybrid machine learning framework that integrates robust variable selection with advanced interaction detection. Specifically, Adaptive LASSO is employed to identify a sparse and interpretable subset of influential features, followed by Bayesian Additive Regression Trees (BART) to model nonlinear effects and uncover key interactions. Selected interactions are further characterized by physiological knowledge through differential equation-informed interaction terms, grounding the model in biologically plausible mechanisms. The insights gained from these steps are then integrated into a final logistic regression model within a Bayesian framework, providing a balance between predictive accuracy and clinical interpretability. This proposed framework not only enhances prediction but also yields actionable insights, offering a valuable support tool for medical research and decision-making.
\vspace{0.15 in}


\end{abstract}
\end{adjustwidth}

\renewcommand{\thefootnote}{\fnsymbol{footnote}} % Use symbols (*, †)
\footnotetext{\hspace{-0.5em}$^{1}$Beijing Key Laboratory of Topological Statistics and Applications for Complex Systems, Beijing Institute of Mathematical Sciences and Applications, Beijing 101408, China.}
\footnotetext{\hspace{-0.5em}$^{2}$Mathematics Department, École Centrale School of Engineering, Mahindra University, Hyderabad, Telangana 500043, India. }
\footnotetext{\hspace{-0.5em}Email: $^{*}$\hspace{-0.5em} \texttt{chitradipachakraborty@gmail.com}, $^{\dagger}$\hspace{-0.5em} \texttt{nayana.mukherjee@mahindrauniversity.edu.in} }

\setcounter{footnote}{0} % Reset for rest of document
\renewcommand{\thefootnote}{\arabic{footnote}} % Restore normal numbers
\section{Introduction}

Gallstone disease is a prevalent gastrointestinal and multifactorial condition characterized by the formation of crystalline deposits, primarily composed of cholesterol or bilirubin, within the gallbladder. These deposits, known as gallstones, form due to an imbalance in the composition of bile, particularly involving cholesterol saturation, bile salts, and phospholipids which can lead to nucleation and crystal aggregation.  Although many individuals with gallstones remain asymptomatic, approximately 20\% develop clinical symptoms such as biliary colic, nausea, and digestive discomfort. In more severe cases, gallstones can migrate and obstruct the biliary or pancreatic ducts, leading to acute conditions such as cholecystitis, cholangitis, or pancreatitis. Long-standing gallstone disease has also been associated with an increased risk of gallbladder carcinoma. Given its global prevalence, with rates between 10\% and 20\% in Western countries (Stinton and Shaffer, 2012; Shaffer, 2005), and its associated economic burden on healthcare systems, early identification of at-risk individuals is critical. Higher prevalence is observed among women, individuals over 40 years of age, and those with obesity, metabolic syndrome, or a sedentary lifestyle. Understanding the multifactorial mechanisms and complex interplay of demographic, clinical, anthropometric, and biochemical factors underlying gallstone formation is therefore essential for guiding effective prevention strategies, diagnosis protocols, and therapeutic decision-making.\\ 
Despite the growing use of data-driven approaches, modeling gallstone disease remains a challenge due to its heterogeneous etiology and intricate relationships among risk factors such as age, sex, body mass index, lipid profiles, and liver function markers. Many of these variables do not operate in isolation but interact in nonlinear and sometimes synergistic ways to influence disease onset. While many cases remain asymptomatic, gallstone disease can progress to serious complications such as cholecystitis, biliary obstruction, pancreatitis, and even gallbladder cancer, particularly when diagnosis or intervention is delayed. The severity and variability of clinical outcomes underscore the importance of accurate risk stratification and early detection. Capturing these interactions requires not only flexible statistical tools but also biologically informed frameworks that reflect the underlying physiology of gallbladder function, bile composition, and metabolic regulation. Therefore, integrating statistical learning methods with mechanistic insights offers a promising direction for improving the interpretability and predictive performance of gallstone disease models.\\
Traditional statistical methods, particularly logistic regression, which is a widely used generalized linear model, have been commonly employed to study the association between risk factors and gallstone status. However, these models often face limitations when dealing with high-dimensional data and complex interactions among predictors (Babyak, 2004). In recent years, advanced statistical and machine learning techniques have shown promise in addressing these challenges by enabling more accurate predictions and uncovering hidden structures in data (Kononenko, 2001; Rajkomar et al., 2019). \\
This study aims to develop a hybrid modeling framework that combines the strengths of variable selection, interaction detection, and mechanistic reasoning to identify key predictors of gallstone status. Specifically, we employ Adaptive LASSO for selecting a sparse and interpretable set of variables (Zou, 2006), followed by Bayesian Additive Regression Trees (BART) to capture nonlinear effects and potential interactions among selected features (Chipman et al., 2010). These interactions, characterized by underlying physiological mechanisms and modeled through a system of ordinary differential equations (ODEs), are subsequently incorporated into a differential equation-informed logistic regression model (Edelstein-Keshet, 2005). This final model is estimated under a Bayesian framework, balancing predictive accuracy with clinical interpretability while embedding mechanistic understanding into the statistical learning process.\\
The analysis is based on data collected from 319 individuals, of whom 158 had gallstones and 161 served as healthy controls. The clinical dataset was collected from the Internal Medicine Outpatient Clinic of Ankara VM Medical Park Hospital and is publicly available through the \texttt{UCI Machine Learning Repository}. The data was introduced and described in detail by Esen et al. (2024) in their study on early prediction of gallstone disease using machine learning techniques. The dataset includes one binary response variable, Gallstone Status, and 38 features encompassing demographic, bioimpedance, and laboratory variables. The study was ethically approved by the Ethics Committee of the Ankara City Hospital Medicine (E2-23-4632).

\noindent \textbf{Demographic Variables:}
Age, Gender, Height, Weight, Body Mass Index (BMI), Comorbidity, Coronary Artery Disease (CAD), Hypothyroidism, Hyperlipidemia, Diabetes Mellitus (DM), and Obesity.

\noindent \textbf{Bioimpedance Measures:}
Total Body Water (TBW), Extracellular Water (ECW), Intracellular Water (ICW), Extracellular Fluid (ECF), Total Body Fat Ratio (TBFR), Lean Mass (LM), Body Protein Content (Protein), Visceral Fat Rating (VFR), Bone Mass (BM), Muscle Mass (MM), Total Fat Content (TFC), Visceral Fat Area (VFA), Visceral Muscle Area (VMA), and Hepatic Fat Accumulation (HFA).

\noindent \textbf{Laboratory Features:}
Glucose, Total Cholesterol (TC), Low Density Lipoprotein (LDL), High Density Lipoprotein (HDL), Triglyceride, Aspartate Aminotransferase (AST), Alanine Aminotransferase (ALT), Alkaline Phosphatase (ALP), Creatinine, Glomerular Filtration Rate (GFR), C-Reactive Protein (CRP), Hemoglobin (HGB), and Vitamin D.\\
By leveraging this rich dataset, our goal is to identify the most influential factors and their interactions that contribute to gallstone formation. The proposed hybrid approach not only improves predictive performance but also yields clinically meaningful interpretations, thereby offering valuable insights for medical research and decision-making.\\
The remainder of the paper is organized as follows. Section 2 outlines the methodological framework adopted in this study, beginning with the use of Adaptive LASSO for efficient variable selection, followed by the application of BART to evaluate feature importance and uncover key interaction effects. These insights are then synthesized into a differential equation-informed logistic regression model, which incorporates both selected predictors and physiologically motivated interactions within a Bayesian framework. Section 3 presents the results, including the variables identified as most influential, the interactions of clinical relevance, and a comparative assessment of model performance across multiple classifiers. This is followed by a detailed presentation of the final Bayesian logistic regression model, emphasizing its interpretability and alignment with clinical knowledge. Lastly, Section 4 discusses the key findings in the context of clinical relevance and previous literature, highlights the advantages and limitations of the proposed approach, and concludes with potential directions for future research.

\section{Methodology}

We employed a multi-stage modeling strategy to balance interpretability, and computational efficiency, and to capture complex patterns. While BART is highly effective at modeling nonlinear relationships and uncovering intricate interactions, it can become computationally intensive in high-dimensional settings. Pre-screening with Adaptive LASSO reduces dimensionality, enabling BART to concentrate on a refined and relevant subset of variables.
This integrated approach leverages the complementary strengths of both methods: the statistical rigor and sparsity-inducing nature of Adaptive LASSO, combined with BART’s flexibility in modeling nonlinearities and detecting interactions. Despite BART's built-in capacity for variable selection, its interpretability and computational efficiency benefit substantially from prior dimensionality reduction via Adaptive LASSO. Early removal of noisy or irrelevant features enhances the robustness and stability of BART’s posterior estimates, leading to clearer insights into variable importance and interaction structures. Importantly, we incorporated these interactions motivated by physiological dynamics and informed by simplified differential equation models. These differential equation-informed interactions reflect biologically plausible relationships and were included to ground the model in mechanistic understanding. Together, this multi-stage modeling framework yields a concise and interpretable logistic regression model that not only enhances predictive performance but also integrates mechanistic insights for greater clinical relevance.

\subsection{Feature Selection with Adaptive LASSO }
In the first stage, we applied the Adaptive Least Absolute Shrinkage and Selection Operator (Adaptive LASSO) to identify a subset of important predictors associated with gallstone status. Originally introduced by Tibshirani (1996), LASSO performs simultaneous variable selection and regularization by imposing an $\ell_1$-penalty on regression coefficients. However, as noted by Fan and Li (2001), standard LASSO tends to produce biased estimates for large coefficients and lacks the oracle property.\\
To address this issue, Zou (2006) proposed the Adaptive LASSO, which introduces data-driven weights into the penalty function: $$\hat{\bm{\beta}}_{AL}=\text{arg min}_{\bm{\beta}} (\bm{y-X\beta})'(\bm{y-X\beta}) + \lambda\sum_{j=i}^p \hat{\omega}_j|\beta_j|,$$ where the weights are defined as $\hat{\omega}=1/|\hat{\beta}|^\gamma$, and $\hat{\beta}$ is a $\sqrt{n}$ consistent estimator with $\gamma > 0$. Under suitable regularity conditions, this formulation enjoys the oracle property and is particularly effective in high-dimensional settings. The Adaptive LASSO simultaneously achieves variable selection and regularization, promoting a sparse solution that improves model interpretability and reduces overfitting. The features with non-zero coefficients in the final Adaptive LASSO model were considered as important predictors and were passed on to the second stage of modeling with BART.

\subsection{Variable Importance and Interaction Detection with BART}
In the second stage, we applied BART, a Bayesian nonparametric ensemble
machine learning method applicable to both regression and classification problems introduced by Chipman et al. (2010). BART builds a sum-of-trees model to capture complex nonlinear relationships and variable interactions. The BART model is expressed as: $$y=\displaystyle\sum_{j=1}^m g(\bm{x}; T_j, M_j) + \epsilon,$$ where each $g(\bm{x}, T_j, M_j)$ represents a regression tree \(T_j\) with associated terminal node values $M_j$ and \(\epsilon\) denotes the random error. Each tree contributes a small part to the overall fit, and the ensemble collectively captures the rich structure in the data. As a Bayesian model, BART adopts a set of prior distributions for the tree structure, terminal node parameters, and residual variance, and posterior inference is conducted via Markov Chain Monte Carlo (MCMC). \\
To assess variable influence, BART uses a method called the inclusion proportion, defined as the proportion of splitting rule for a given variable across the posterior samples of the tree structures. Let $c_{rk}$ be the number of splitting rules using the $r$th predictor as a split variable in the $k$th posterior sample of the trees’ structure across $m$ trees, and
$c_{.k}=\displaystyle\sum_{r=1}^p c_{rk}$ represent the total number of splitting rules across the total $p$ variables. Therefore, the variable importance score of $r$th variable is given by $$\text{Vimp}_r=\frac{1}{K}\displaystyle\sum_{k=1}^K \frac{c_{rk}}{c_{.k}}.$$ 
To detect interactions, we adopted the interaction score by observing successive splitting rules in each tree (Kapelner and Bleich, 2016). Let $c_{rqk}$ be the number of splitting rules using predictors $r$ and $q$ successively (in either order) in the $k$th posterior sample, and $c_{..k}=\displaystyle\sum_{r=1}^p \sum_{q=1}^p c_{rqk}$ represent the total number of successive splitting rules. Therefore, the interaction score between variables $r$ and $q$ is $$\text{Vint}_{rq}=\frac{1}{K}\displaystyle\sum_{k=1}^K \frac{c_{rqk}}{c_{..k}}.$$
To visualize these results, we used a heatmap approach proposed by Inglis et al. (2022) to display both importance and interactions simultaneously.  Since both the Vimp and Vint values are calculated from the full posterior, it is trivial to compute an uncertainty associated with their measurement. To represent posterior uncertainty, we employed Value Suppressing Uncertainty Palettes (VSUP) heatmap, introduced by Inglis et al. (2024), which encodes uncertainty alongside magnitude to suppress misleading interpretations. In both heatmaps, the variable importance is displayed on the diagonal and the interactions are on the off-diagonal. These visualizations allow intuitive exploration of both main effects and interactions derived from the full BART posterior.

\subsection{Logistic Regression with Differential Equation-Informed Interactions}

In the final stage of our multi-stage modeling framework, we employed a logistic regression model to estimate the probability of gallstone disease (Christodoulou et al., 2019). Logistic regression models the log-odds of a binary outcome as a linear combination of predictor variables and their interactions, providing interpretable coefficient estimates that facilitate clinical insight. Following pre-screening via Adaptive LASSO, the model incorporated a refined subset of predictors and interaction terms identified by BART, prioritizing those with the strongest evidence of nonlinearity and synergy. Notably, these interactions were further characterized by domain-specific knowledge and underlying physiological mechanisms, modeled as differential equation-informed interaction terms. These terms represent biologically plausible, dynamic relationships between variables, capturing how one physiological quantity may influence the rate of change of another (Kitano, 2002). This hybrid modeling framework, blending data-driven interaction discovery with mechanistic insight, ensures that the final model remains parsimonious and clinically interpretable while capturing essential non-linear dependencies. The final model is expressed as

\[
\log \left(
\frac{P(Y_i = 1)}{P(Y_i = 0)}
\right)
= \beta_0 + \sum_{j=1}^p \beta_j X_{ij} + \sum_{(j,k) \in I} \gamma_{jk} \left( X_{ij} \times X_{ik} \right),
\]

\noindent where \(X_{ij}\) denotes the \(j\)-th feature for observation \(i\), \(I\) is the set of index pairs \((j,k)\) for selected interactions, and \(\beta_0, \beta_j\) and \(\gamma_{jk}\) represent the intercept, main effect coefficients and interaction coefficients, respectively. Each pair was modeled using a first-order differential equation, where the rate of change of one variable with respect to another reflects not only the nature of their relationship but also how the interaction, denoted by ``\(\times\)'', evolves in a non-linear fashion (Keener and Sneyd, 2009; Murray, 2002). The specific functional form of each differential equation was constructed to reflect both the underlying theoretical interaction structure and the trends observed in the data. These interactions were derived from observed data and represent meaningful physiological relationships between selected variable pairs (Brauer and Castillo-Chavez, 2011; Voit, 2012). By modeling interactions in this way, we capture how physiological variables co-evolve in a non-linear and biologically plausible manner, providing insight into their mutual dependencies. The model was estimated within a Bayesian framework to incorporate prior uncertainty and provide robust posterior inference (Gelman et al., 2013).

\section{Analysis and Results}
In this section, we present a comprehensive analysis combining variable selection, interaction detection, and predictive modeling to identify key predictors and their synergistic effects related to the binary outcome. The approach integrates Adaptive LASSO for feature selection, BART for uncovering interactions, and a comparison of three classification models: BART, Logistic Regression, and Random Forest (Breiman, 2001). Leveraging insights from the variable selection and interaction analyses, we construct a final logistic regression model incorporating the most influential predictors, significant interaction terms identified by BART, and differential equation-informed interactions grounded in physiological mechanisms. This hybrid model balances predictive accuracy with interpretability, providing clear effect estimates consistent with the clinical understanding of gallstone risk factors.

\subsection{Selected Features and Interaction Insights}

The Adaptive LASSO penalty was implemented using the \texttt{glmnet} package in R (Friedman et al., 2010), with the penalty weights derived from an initial ordinary least square (OLS) estimates. To select the optimal penalty parameter ($\lambda$), we used 10-fold cross-validation to \\

\begin{figure}[htbp]
    \centering
        \includegraphics[width=0.9\linewidth]{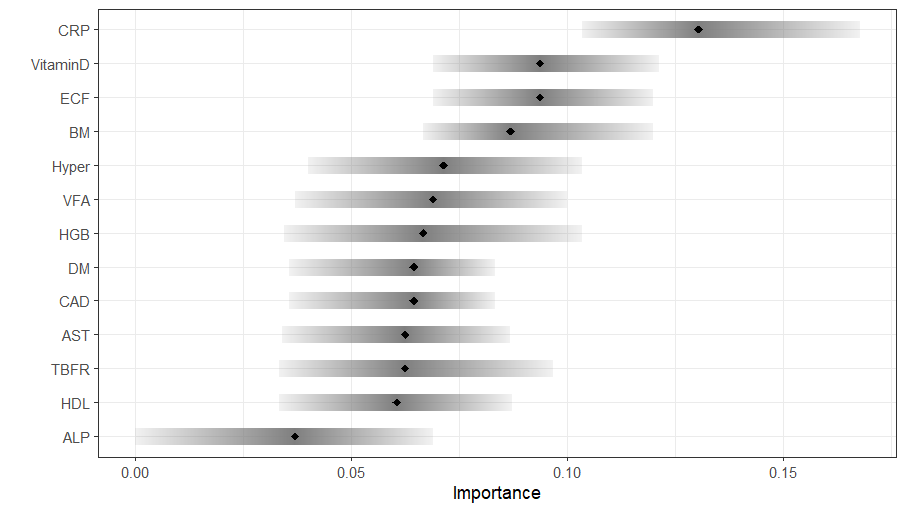}
        \caption{Inclusion Proportions with the 25\% to 75\% Quantile Interval}
        \label{fig:inc_prop}
\end{figure}  

\noindent minimize prediction error. This approach reduced the initial set of 38 features to 13 key predictors, promoting model parsimony and interpretability, critical aspects in biomedical research. The 13 features selected by Adaptive LASSO were: CAD, Hyperlipidemia, DM, ECF, TBFR, BM, VFA, HDL, AST, ALP, CRP, HGB, and Vitamin D. From this subset, BART identified the following nine variables as most important (Figure 1): CRP, Vitamin D, ECF, BM, Hyperlipidemia, VFA, HGB, DM, and CAD. BART was implemented using the \texttt{bartMachine} package in R  (Kapelner and Bleich, 2016) with default settings of 1000 iterations with a burn-in of 250, and with the number of trees to be 20.\\

\begin{figure}[htbp]
  \centering
  \begin{minipage}[b]{0.49\textwidth}
    \centering
    \includegraphics[width=\textwidth]{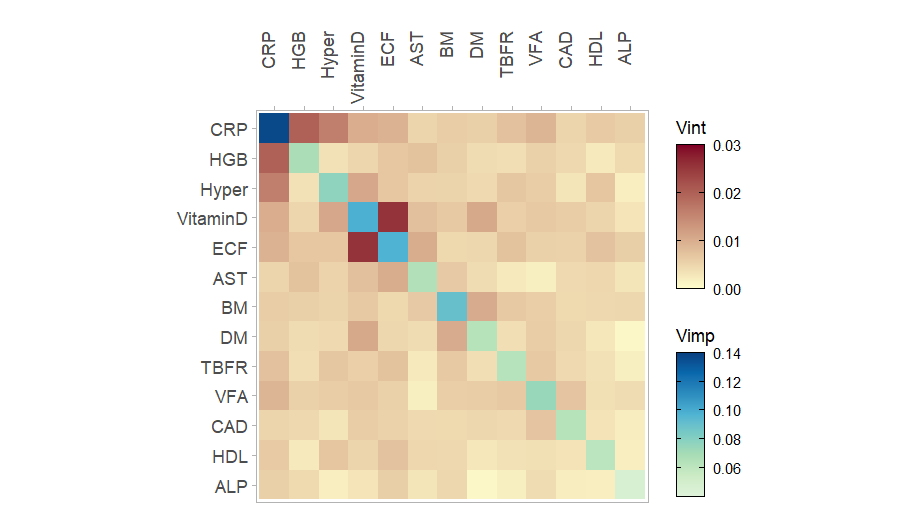}
    \subcaption{}
    \label{fig:sub_heatmap}
  \end{minipage}
  \hfill
  \begin{minipage}[b]{0.49\textwidth}
    \centering
    \includegraphics[width=\textwidth]{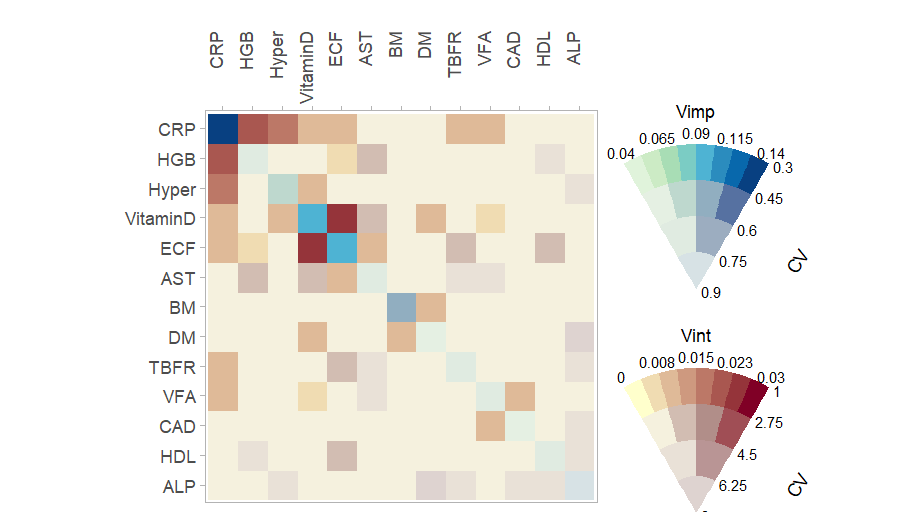}
    \subcaption{}
    \label{fig:sub_vsup}
  \end{minipage}
  \caption{(a) Heatmap of Interaction Strengths (Vimp and Vint); (b) VSUP-adjusted Heatmap with Coefficient of Variation (CV)}
  \label{fig:interaction_heatmap_comparison}
\end{figure}

\noindent To explore potential feature interactions, we analyzed two heatmaps. Figure 2(a) presents the interaction strengths using BART's variable inclusion proportions, while Figure 2(b) shows the VSUP-adjusted heatmap, incorporating the Coefficient of Variation (CV) to assess interaction stability (Bien et al., 2013). These analyses highlighted several notable interactions, including a strong and highly stable interaction between Vitamin D and ECF, as well as a strong and stable interaction between CRP and HGB. Additionally, Vitamin D and Hyperlipidemia showed a strong and stable interaction, while BM and DM exhibited a moderate to strong interaction with good stability. These interactions suggest potential synergistic effects among clinical variables relevant to metabolic and inflammatory processes. 

\subsection{Model Comparison}

In this section, we evaluated the predictive performance of three models: BART, Logistic Regression, and Random Forest, using Precision (Positive Predictive Value), Recall (Sensitivity), AUC (Area Under the ROC Curve), and F1 Score as evaluation metrics with the important features and interaction terms identified via Adaptive LASSO and BART (Fawcett, 2006; Powers, 2011; Hastie et al., 2009; Murphy, 2012). As shown in Table 1, BART achieved the highest AUC (0.9316), Recall (0.8820), and F1 Score (0.8529), followed by logistic regression. While logistic regression correctly identified more true positives, it\\

\begin{table}[ht]
\centering
\begin{tabular}{lcccc}
\hline
\textbf{Model} & \textbf{Precision} & \textbf{Recall} & \textbf{AUC} & \textbf{F1 Score} \\
\hline
BART & 0.8256 & 0.8820 & 0.9316 & 0.8529 \\
Logistic Regression & 0.7797 & 0.8571 & 0.8937 & 0.8166\\
Random Forest & 0.7840 & 0.7950 & 0.8533 &  0.7895\\
\hline
\end{tabular}
\caption{Comparison of Model Performance}
\end{table}
\noindent also incurred a higher rate of false positives. In contrast, Random Forest was more conservative in predicting positives with high precision but it missed more actual positives, resulting in the lowest recall among the three. This comparison highlights the superior predictive ability of BART on our dataset. Figure 3 illustrates the comparative accuracy of the three models. BART consistently outperforms both Random Forest and Logistic Regression, further reinforcing its robustness in prediction.\\
Although Random Forest is a widely used and reliable ensemble machine learning method that enhances accuracy and reduces overfitting by aggregating multiple decision trees, its relatively low recall, AUC, and F1 Score indicate poor performance in this context. It failed to adequately identify true positives and to maintain a strong balance between precision and recall, both of which are critical to our clinical prediction goal. Additionally, the model\\
\begin{figure}[htbp]
    \centering
        \includegraphics[width=0.9\linewidth]{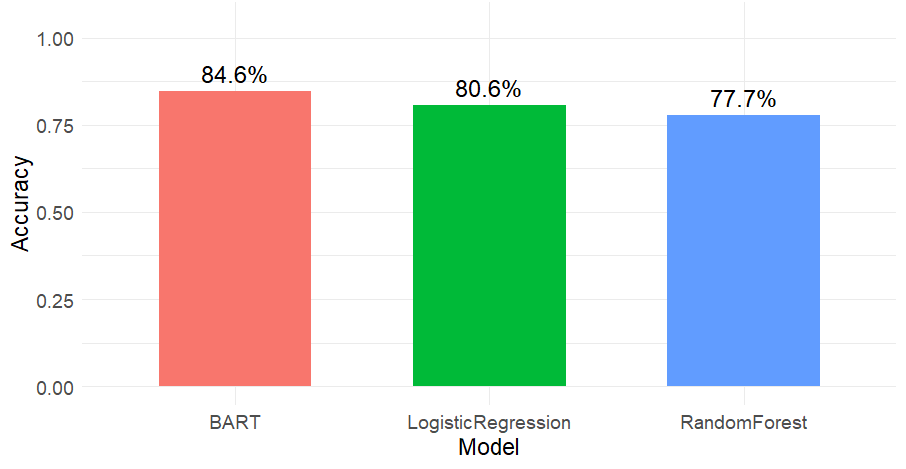}
        \caption{Comparative Accuracy}
        \label{fig:heatmap}
\end{figure}

\noindent exhibited low overall accuracy in predicting the risk of gallstone, further limiting its utility for this application. This relatively weaker performance supports the selection of BART and Logistic Regression for variable selection and model development. Including Random Forest in the comparison provides a benchmark against a standard machine learning model, which adds rigor and comprehensiveness to our study.\\ 
While BART demonstrated strong predictive performance, it does not provide explicit coefficient estimates, which limits its interpretability. Logistic regression, on the other hand, produces interpretable coefficients that quantify both the direction and magnitude of associations between predictors and the binary response (Hosmer et al., 2013). To leverage the strengths of both methods, we adopted the hybrid modeling strategy. First, BART was used to identify a parsimonious set of important features and interactions. These interactions were further characterized by physiological knowledge and modeled as differential equation-informed interaction terms, reflecting biologically plausible dynamic relationships. Then, a logistic regression model was fitted in a Bayesian setting using this refined set of predictors and interactions to obtain interpretable coefficient estimates.\\
This multi-stage approach combines the variable selection and interaction detection capabilities of BART with the interpretability and mechanistic grounding of logistic regression. It facilitates both accurate prediction and meaningful scientific inference. In summary, this modeling framework ensures that the final model is predictively strong, scientifically transparent, and grounded in clinical relevance, aligning with our objectives of achieving high accuracy and interpretability.

\subsection{Final Bayesian Logistic Regression with Interaction Effects}

The nonlinear interaction dynamics, modeled using differential equations, is illustrated in Figure 4. These interactions are derived from observed data and reflect meaningful physiological relationships between selected pairs of variables. The governing differential equations are:
\begin{align}
\frac{d(\text{ECF})}{d(\text{VitD})} &= \text{ECF} \cdot \log(1 + \text{VitD}) \tag{1} \\
\frac{d(\text{CRP})}{d(\text{HGB})} &= \frac{\text{CRP}}{1 + \text{HGB}} \tag{2} \\
\frac{d(\text{Hyper})}{d(\text{VitD})} &= -2 \cdot \text{VitD} \cdot \text{Hyper} \tag{3} \\
\frac{d(\text{DM})}{d(\text{BM})} &= \text{BM}^2 \cdot \text{DM} \tag{4}
\end{align}

\noindent Equation (1) describes the relationship between ECF and Vitamin D (VitD). As shown in Figure 4(a), the model suggests that Vitamin D contributes to elevated ECF levels, though in a moderated manner. This equation indicates that ECF increases with Vitamin D, but the rate of increase diminishes over time. The logarithmic form captures a saturating growth pattern, where Vitamin D stimulates ECF accumulation more prominently at lower\\

\begin{figure}[htbp]
  \centering

  % Row 1
  \begin{minipage}[b]{0.48\linewidth}
    \centering
    \includegraphics[width=\linewidth]{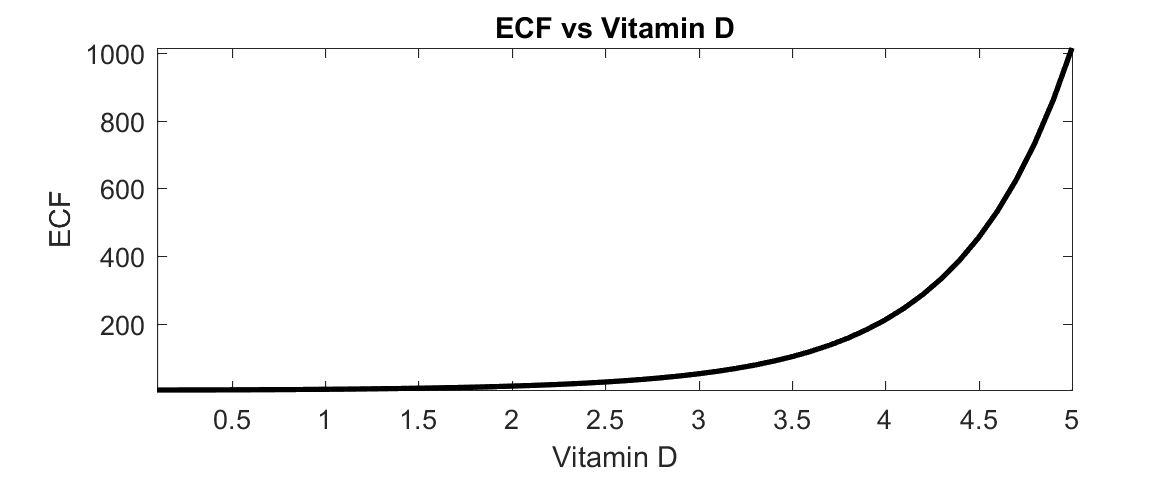}
    \subcaption{}
    \label{fig:sub1}
  \end{minipage}
  \hfill
  \begin{minipage}[b]{0.48\linewidth}
    \centering
    \includegraphics[width=\linewidth]{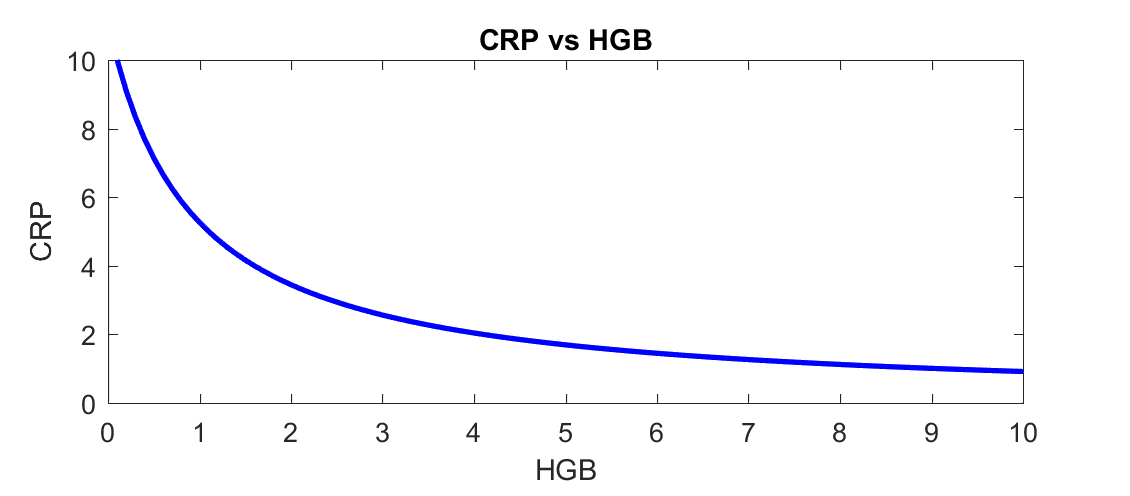}
    \subcaption{}
    \label{fig:sub2}
  \end{minipage}

  \vspace{0.5cm}

  % Row 2
  \begin{minipage}[b]{0.48\linewidth}
    \centering
    \includegraphics[width=\linewidth]{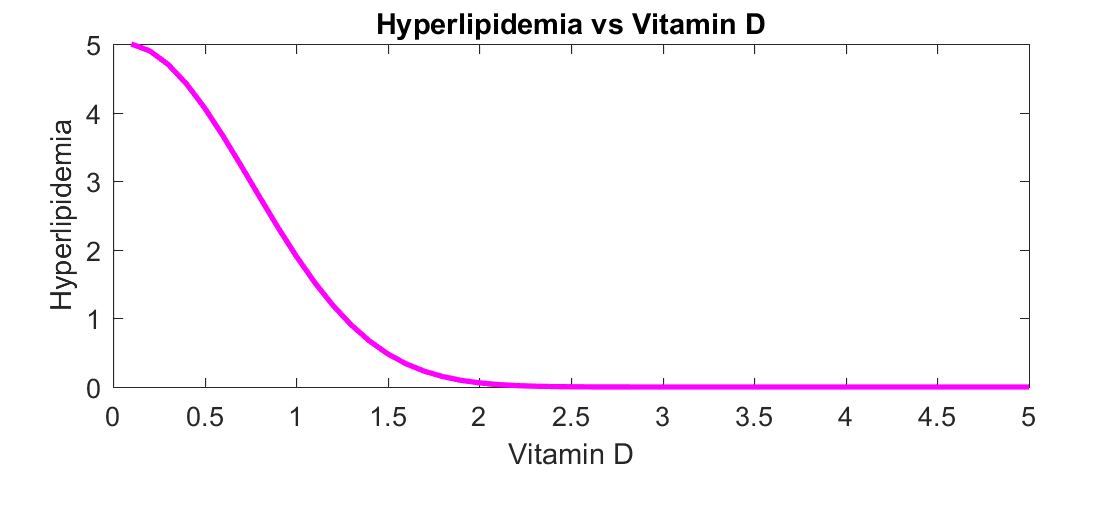}
    \subcaption{}
    \label{fig:sub3}
  \end{minipage}
  \hfill
  \begin{minipage}[b]{0.48\linewidth}
    \centering
    \includegraphics[width=\linewidth]{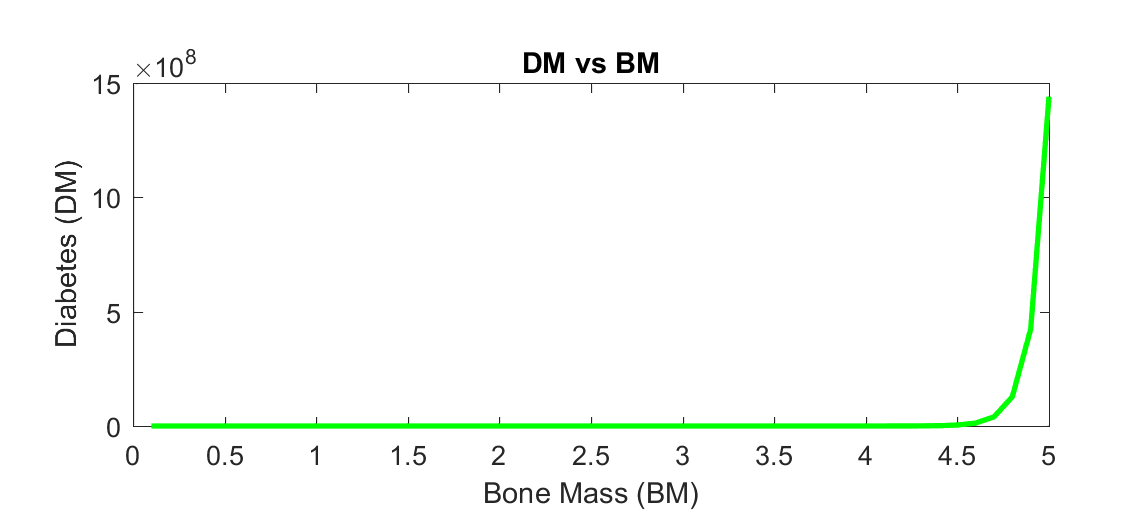}
    \subcaption{}
    \label{fig:sub4}
  \end{minipage}

  \caption{Simulated plots of differential interaction models between selected physiological variables}
  \label{fig:interaction_heatmaps_all}

\end{figure}

\noindent  levels, while the effect weakens as Vitamin D continues to rise. Equation (2) captures the interaction between CRP and HGB, shown in Figure 4(b). This equation indicates that CRP decreases with increasing HGB, with the rate of decline tapering off at higher HGB levels. The rate of decrease is proportional to the current CRP level and inversely dependent on the hemoglobin concentration. This structure reflects a saturating inhibitory effect where initial increases in HGB lead to a substantial reduction in CRP, although the magnitude of this effect diminishes at higher HGB levels. Equation (3), as shown in Figure 4(c), models the relationship between Vitamin D and Hyperlipidemia (Hyper), reflecting that an increase in Vitamin D may be associated with a decrease in hyperlipidemia. This formulation suggests a multiplicative decay, where the rate of decline in hyperlipidemia is directly proportional to both the current Vitamin D level and the existing level of hyperlipidemia. The equation represents a feedback mechanism in which higher Vitamin D levels and elevated lipid risk jointly accelerate the decline of that risk. Equation (4), illustrated in Figure 4(d), suggests that the likelihood of DM increases slightly as BM increases. This equation implies that the rate of increase in diabetes is influenced not only by the current presence of diabetes but also by the square of the bone mass. In other words, higher bone mass intensifies the impact of existing diabetic conditions, reflecting a nonlinear compounding effect.\\
Therefore, the final logistic regression model is expressed as:
\[
\begin{aligned}
\log \left( \frac{P(Y_i=1)}{P(Y_i=0)} \right) = \ & \beta_0 
+ \beta_1 \cdot \text{CRP}_i + \beta_2 \cdot \text{VitD}_i + \beta_3 \cdot \text{ECF}_i + \beta_4 \cdot \text{BM}_i \\
& + \beta_5 \cdot \text{Hyper}_i + \beta_6 \cdot \text{VFA}_i + \beta_7 \cdot \text{HGB}_i + \beta_8 \cdot \text{DM}_i + \beta_9 \cdot \text{CAD}_i \\
& + \gamma_1 \cdot \left( \text{ECF}_i \cdot \log(1 + \text{VitD}_i) \right) + \gamma_2 \cdot \left( \frac{\text{CRP}_i}{1 + \text{HGB}_i} \right) \\
& + \gamma_3 \cdot \left( -2 \cdot \text{VitD}_i \cdot \text{Hyper}_i \right) + \gamma_4 \cdot \left( \text{BM}_i^2 \cdot \text{DM}_i \right),
\end{aligned}
\]

\noindent where \(\beta_0\) denotes the intercept; \(\beta_j\) for \(j = 1, \ldots, 9\) represent the coefficients of the main effects; and \(\gamma_k\) for \(k = 1, \ldots, 4\) correspond to the coefficients of the interaction terms.
To estimate the model within a Bayesian framework, we used the \texttt{brm()} function from the \texttt{brms} package in R (Bürkner, 2017). For the regression coefficients, we used weakly informative normal priors, specifically $N(0,5)$ for all fixed effects and $N(0,10)$ for the intercept (Gelman et al., 2008). The model was estimated using MCMC with 4 chains, each run for 4000 iterations, including a warm-up of 1000 iterations per chain. Convergence was assessed using standard diagnostics, such as trace plots and the Gelman-Rubin statistic ($\hat{R}$), with all parameters achieving $\hat{R}$ values close to 1, indicating good convergence (Gelman and Rubin, 1992). \\
The results presented in Table 2 exhibit strong alignment with existing clinical literature on gallstone risk factors. Elevated CRP is strongly associated with increased gallstone risk, highlighting the central role of inflammation in gallstone formation (Maurer et al., 2009; Rajab et al., 2020). VFA also contributes positively, consistent with the impact of central\\

\begin{table}[ht]
\centering
\label{tab:posterior-summary}
\begin{tabular}{lrrrrrrrr}
\hline
\textbf{Parameter} & \textbf{Estimate} & \textbf{Est. Error} & \multicolumn{2}{c}{\textbf{95\% CI}} & \textbf{Rhat} &  \multicolumn{2}{c}{\textbf{ESS}} \\
 \textbf{(Coefficient)}               &                   &                      & \textbf{Lower} & \textbf{Upper}      &                &     \textbf{Bulk}               &        \textbf{Tail}           \\
\hline
Intercept (\(\beta_0\))         & 21.11  & 4.53  & 12.86  & 30.54  & 1.00 & 8064  & 8593  \\
CRP (\(\beta_1\))               & 1.92   & 0.32  & 1.30   & 2.58   & 1.00 & 5603  & 6633  \\
VitD (\(\beta_2\))          & -0.10  & 0.07  & -0.24  & 0.03   & 1.00 & 5257  & 5741  \\
ECF (\(\beta_3\))               & -0.35  & 0.12  & -0.59  & -0.12  & 1.00 & 5550  & 6286  \\
BM (\(\beta_4\))                & -1.79  & 0.49  & -2.78  & -0.87  & 1.00 & 7682  & 7548  \\
Hyper (\(\beta_5\))             & 0.23   & 4.87  & -9.17  & 9.92   & 1.00 & 10686 & 7021  \\
VFA (\(\beta_6\))              & 0.15   & 0.04  & 0.07   & 0.24   & 1.00 & 8566  & 7981  \\
HGB (\(\beta_7\))               & -0.21  & 0.12  & -0.45  & 0.01   & 1.00 & 9296  & 7594  \\
DM (\(\beta_8\))                & 0.89   & 1.26  & -1.55  & 3.40   & 1.00 & 5898  & 6250  \\
CAD (\(\beta_9\))               & -1.41  & 0.94  & -3.41  & 0.30   & 1.00 & 10047 & 6832  \\
ECF $\times$ VitD (\(\gamma_1\))   & 0.01   & 0.03  & -0.05  & 0.07   & 1.00 & 5115  & 5724  \\
CRP $\times$ HGB (\(\gamma_2\))    & -17.93 & 3.33  & -24.45 & -11.47 & 1.00 & 5541  & 6409  \\
VitD $\times$ Hyper (\(\gamma_3\))  & -3.54  & 2.57  & -9.54  & -0.22  & 1.00 & 3599  & 3977  \\
BM $\times$ DM (\(\gamma_4\))      & 0.03   & 0.14  & -0.24  & 0.30   & 1.00 & 5862  & 6431  \\
\hline
\end{tabular}
\caption{Posterior Summary Statistics for Model Parameters, where $\times$ denotes interaction}
\end{table}

\noindent obesity on metabolic health. Vitamin D has a small negative association, suggesting that higher Vitamin D levels may slightly reduce the risk of gallstones, though this main effect is modest and not statistically significant. Similarly, increased ECF and higher BM show significant negative associations with gallstone risk, possibly reflecting the role of frailty and fluid imbalance in increasing susceptibility to the disease. Lower HGB levels are linked to higher gallstone risk, reflecting the adverse influence of anemia on metabolic and liver function; however, this association is not statistically significant. While DM, hyperlipidemia, and CAD do not emerge as significant independent predictors, their clinical relevance becomes evident through interaction terms.\\

\begin{figure}[htbp]
  \centering
  \begin{minipage}[b]{0.48\linewidth}
    \centering
    \includegraphics[width=\linewidth]{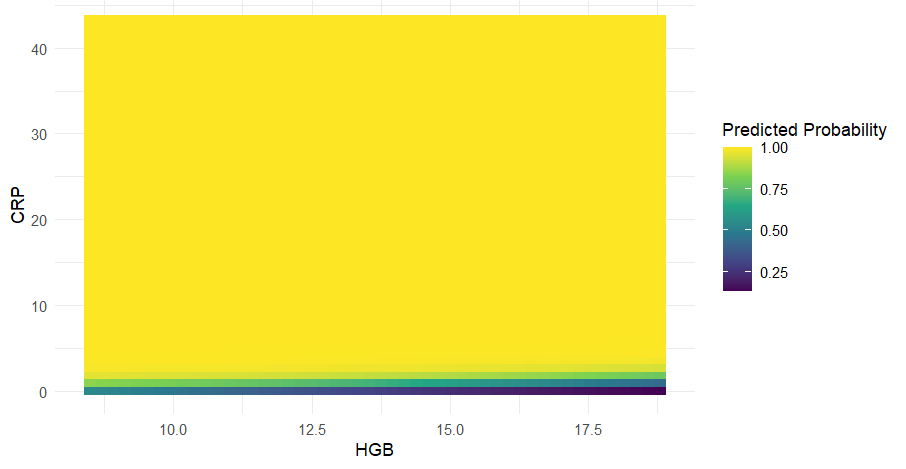}
    \subcaption{}
    \label{fig:sub1}
  \end{minipage}
  \hfill
  \begin{minipage}[b]{0.48\linewidth}
    \centering
    \includegraphics[width=\linewidth]{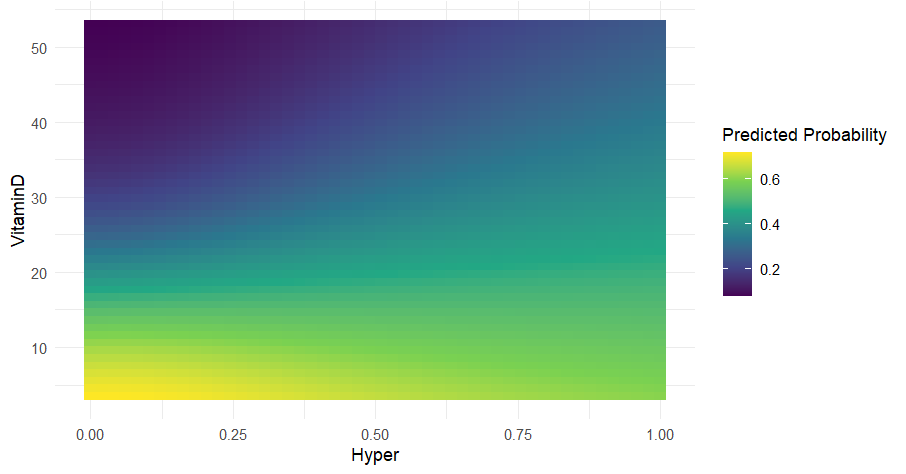}
    \subcaption{}
    \label{fig:sub2}
  \end{minipage}
  \caption{Interaction effects on predicted gallstone risk: (a) C-reactive protein and Hemoglobin; (b) Vitamin D and Hyperlipidemia.}
  \label{fig:interaction_heatmaps}
\end{figure}

\noindent Among these, the interaction between CRP and HGB is strongly negative, indicating that low hemoglobin substantially amplifies gallstone risk in the presence of elevated systemic inflammation. As illustrated in Figure 5(a), predicted probabilities of gallstone risk approach nearly 100\% when CRP is high and hemoglobin is low. This highlights a critical clinical insight that the combination of anemia and systemic inflammation markedly increases the likelihood of gallstone development. Additionally, Vitamin D interacts negatively with hyperlipidemia, suggesting that adequate Vitamin D levels may mitigate the increased gallstone risk associated with lipid metabolism disturbances (Ponda et al., 2012). This is consistent with the observed interaction in Figure 5(b), where the predicted risk sharply declines with increasing Vitamin D, particularly in individuals with hyperlipidemia. Other interaction terms, such as ECF with Vitamin D, and BM with DM were not statistically significant. Together, these findings provide meaningful clinical insight into the combined effects of inflammation, metabolic status, and nutritional factors on gallstone risk.

\section{Conclusion}

This study introduces a hybrid Bayesian machine learning framework that combines Adaptive LASSO for variable selection, BART for detecting nonlinear effects and interactions, and a differential equation-informed logistic regression model for clinical interpretability. Applied to a detailed clinical dataset on gallstone disease, this framework offers robust and clinically meaningful insights, highlighting the complex interplay among metabolic, inflammatory, and nutritional factors (Ghorbani, 2023). Our key findings show that elevated CRP, increased VFA, low BM, and reduced ECF are significantly associated with a higher risk of gallstone disease. These results align well with existing clinical literature that highlights the roles of systemic inflammation, central obesity, and physical frailty in gallstone formation (Shabanzadeh, 2017). Importantly, our model reveals interaction effects, especially between CRP and HGB, and between Vitamin D and hyperlipidemia, suggesting that anemia and inflammation, as well as nutritional and lipid imbalances, synergistically increase the likelihood of gallstone development. A major strength of our proposed approach is its ability to balance accuracy and clinical interpretability. Adaptive LASSO helps select only the most relevant predictors, BART captures nonlinear and complex relationships, while the final logistic regression layer, informed by physiological reasoning, maintains transparency for clinical decision-making. This multi-stage design enhances model parsimony and reduces the risk of overfitting, a common concern of high-capacity machine learning models in biomedical settings.\\
However, this study has a few limitations. The analysis is based on a single-center dataset with a moderate sample size (N = 319), which may limit the generalizability of the findings to broader populations, especially those with different demographic, lifestyle, or genetic backgrounds. The clinical data used were cross-sectional, capturing patient information at a single time point, which restricts our ability to make causal inferences. Validation through longitudinal or experimental designs is necessary to establish temporal or causal relationships. Additionally, the use of advanced modeling techniques such as BART and differential equation-informed regression, while powerful, requires careful calibration and interpretive caution in smaller datasets due to potential overfitting and model instability. While the inclusion of physiologically motivated interaction terms adds interpretive value, these interactions should be viewed as hypothesis-generating rather than confirmatory.\\ 
In terms of future directions, several avenues hold promise. First, expanding the study to include larger and more diverse datasets from multi-center cohorts would help validate the robustness and generalizability of the model across various populations and clinical settings. Second, integrating longitudinal data, such as repeated measures of biomarkers, lifestyle behaviors, and disease progression, could offer richer insights into the temporal dynamics and causal mechanisms underlying gallstone disease. Third, translating the model into a real-time clinical decision support tool that leverages electronic health records (EHRs) could make the findings actionable in routine care, facilitating personalized risk stratification and early intervention. Such translational efforts would require further validation, interpretability safeguards, and prospective evaluation within clinical workflows.\\
In summary, our hybrid Bayesian framework not only improves risk prediction but also uncovers meaningful and clinically interpretable interactions among key physiological domains, namely, inflammation, metabolic health, and nutrition, that contribute to gallstone disease. This work illustrates the strength of combining statistical rigor, flexible machine learning methods, and domain-specific knowledge in building predictive and explanatory models for complex diseases.

\end{document}